\documentclass[pre,twocolumn,showpacs]{revtex4}
\usepackage{graphicx}
\usepackage{dcolumn}
\usepackage{amsmath}
\begin{document}
\title[Short Title]{Remarks on $(1-q)$ expansion and factorization
approximation in the Tsallis nonextensive statistical mechanics}
\author{E. K. Lenzi$^{1}$, R. S. Mendes$^{2}$, L. R. da Silva$^{3,4}$,
and L. C. Malacarne$^{2}$}
\affiliation{$^{1}$Centro Brasileiro de Pesquisas F\'{\i}sicas, \\
R. Dr. Xavier Sigaud 150, 22290-180 Rio de Janeiro-RJ, Brazil\\
$^2$Departamento de F\'{\i}sica, Universidade Estadual de Maring\'a, \\
Av. Colombo 5790, 87020-900 Maring\'a-PR, Brazil \\
$^3$Center for Polymer Studies, Boston University, Boston, Ma 02215, USA \\
$^4$ Departamento de F\'{\i}sica, Universidade Federal do Rio Grande do\\
Norte, 59072-970 Natal-RN, Brazil }
\date{\today}

\begin{abstract}
The validity of (1-q) expansion and factorization approximations
are analysed in the framework of Tsallis statistics. We employ
exact expressions for classical independent systems (harmonic
oscillators) by considering the unnormalized and normalized
constrainsts. We show that these approxiamtions can not be
accurate in the analysis of systems with many degrees of freedom.

\end{abstract}

 \maketitle


\section*{I-Introduction}

Ever since the presentation by Tsallis\cite{ts88,ct91,NORMA} of a
new possible generalization of the statistical mechanics (Tsallis
statistics) based on the nonextensive entropy,
\begin{eqnarray}
S_{q}=k\frac{1-{\mbox {Tr}}\rho _{q}^{q}}{q-1} \;
,\label{entropia}
\end{eqnarray}
a large number of investigations has been developed concerning
this subject \cite{web}. These investigations are basically
employed in the discussion of aspects related to nonextensive
phenomena, such as, L\'{e}vy-type anomalous superdiffusion
\cite{6}, Euler turbulence\cite{7}, self-gravitating systems and
related themes\cite{7,8,8a,8b,8c}, cosmic background radiation
\cite{9,exato,condmat,resto1}, peculiar velocities in
galaxies\cite{10a}, eletron-phonon interaction\cite{10c}, and
ferrofluid-like systems\cite{11}. In addition, some important
methods of the usual Boltzmann-Gibbs statistics
 have been generalized in order to incorporate the Tsallis
 framework, {\it {e. g.}}, linear response theory \cite{10b}, Green function
theory\cite{RAJA} and path integral\cite{PATHINT}. However,  exact
calculations are generally difficult to be performed in the
context of Tsallis statistics. Motivated by these difficulties
some approximated methods, such as $(1-q)$ expansion
\cite{9,resto1,cp97,cu96}, factorization approximation
\cite{mmmm}, perturbative expansion \cite{lm98}, generalized
Bogoliubov inequality \cite{lm98,pt95}, and semi-classical
expansion\cite{em98}, have been developed. The two first methods
mentioned above have been employed largely in the analysis of
black body radiation \cite{9,condmat,resto1} and in other many
independent particle systems
\cite{QA,pennini,outrasreferencias,turco1,IVAN,resto2}, but
without a careful analysis of the validity of these methods. Thus,
a detailed discussion of the $(1-q)$ expansion and the
factorization approximation plays a special role in this scenario.
In this direction, Ref. \cite{DF} contains a discussion comparing
these approximations in the context of the quantum gases. However,
some important aspects of the comparison are still lacking, in
particular the importance of the degrees of freedom $N$. This work
is addressed to analyze how precise the $(1-q)$ expansion and the
factorization approximation are by considering $N$ arbitrary. To
perform this study it is convenient to consider a solvable model
in order to carefully understand the degree of accuracy of these
approximations. Furthermore, the chosen model must contain
important features of other more realistic ones. This is just the
case of a set of harmonic oscillators. In this direction, to
analyze the $(1-q)$ expansion and the factorization approximation,
we focus our discussion on the classical Tsallis statistics by
considering $N$  one dimesional harmonic oscillators, {\it
{i. e.}}, we employ the Hamiltonian
\begin{equation}
H= \sum_{n=1}^{N}\left(\frac{p_{n}^{2}}{2m}+\frac{1}{2}m\omega_n
^{2}x_{n}^{2}\right)\;\;, \label{1}
\end{equation}
where $\frac{p_{n}^{2}}{2m}+\frac{1}{2}m\omega_n
^{2}x_{n}^{2}$ is the Hamiltonian to the $n$-th harmonic
oscillator, being $m$  the particle mass and $\omega_n $  the
harmonic oscillator frequency.

This work is organized as follows. In Sec. II, we investigate both
approximations by using unnormalized constraints. The analysis of
these approximations employing normalized constraints is done in
Sec. III and the conclusions are presented in Sec. IV.

\section*{II-Exact versus approximate calculations and unnormalized constraints}

\begin{figure}
 \centering
 \DeclareGraphicsRule{ps}{eps}{*}{}
 \includegraphics*[width=10cm, height=14cm,trim=0cm 0cm 0cm 0cm]{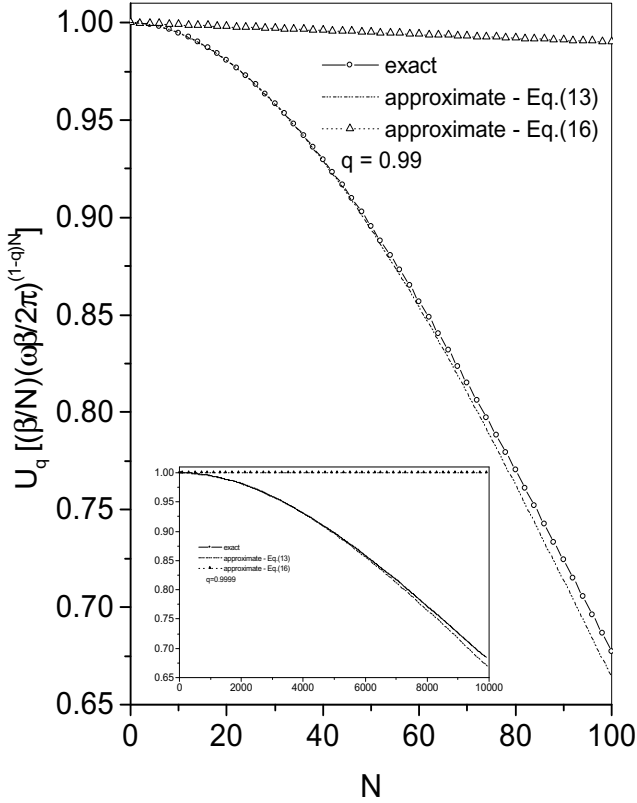}
\caption{In order to illustrate the differences among Eqs.
(\ref{UUqq}), (\ref{expand}) and (\ref{mala}) we ploted them
versus $N$ for $q=0.99$ and $q=0.9999$.}
\label{fig1}
\end{figure}

In the Tsallis statistics with unnormalized constraints,  the
nonextensive canonical distribution is given by
\begin{equation}
\rho _{q}=\frac{1}{Z_{q}}[1-(1-q)\beta H]^{1/(1-q)} \;\;, \label{2}
\end{equation}
where
\begin{eqnarray}
Z_{q}=\int \prod_{n=1}^{N}dp_{n}dx_{n}[1-(1-q)\beta H]^{1/(1-q)}\;
\label{partition}
\end{eqnarray}
is the partition function with $q\in {\mbox {\bf R}}$ and $\beta $
$\equiv $ $1/T$ (with $k=1$). Eq.(\ref{2}) is obtained maximizing
the Tsallis entropy\cite{ts88,ct91}, Eq. (\ref{entropia}), subject
to the constraints
\begin{eqnarray}
U_{q}=\int \prod_{n=1}^{N}{\mbox d}p_{n}{\mbox d}x_{n}\;\rho
_{q}^{q}\;H\; \label{Uq}
\end{eqnarray}
and
\begin{eqnarray}
\int \prod_{n=1}^{N}{\mbox d}p_{n}{\mbox d}x_{n}\;\rho _{q}=1\; ,
\end{eqnarray}
with $U_q$ being the generalized internal energy. In this context,
it is employed
\begin{equation}
A_{q}=\langle A \rangle_q =\int \prod_{n=1}^{N}{\mbox d}p_{n}{\mbox d}x_{n}\;\rho
_{q}^{q}\;A
\label{VINCULO1}
\end{equation}
as the generalized mean value of the classical function $A(p,x)$
and thermodynamical functions like free energy and specific heat
are defined as $ F_{q}=(Z_{q}^{q-1}-1)/[(1-q)\beta ]\;,$ and
$C_{q}=\partial U_{q}/\partial T$. Furthermore, the Legendre
structure of the Tsallis statistics is preserved
\cite{ct91,pp97,me97}; thus, $F_{q}=U_{q}-T\;S_{q}$,
$S_{q}=-\partial F_{q}/\partial T$, etc. In Eq. (\ref{2}) we
assumed that $1-\left( 1-q\right) \beta H\geq 0$. When this
condition is not satisfied we have a cut-off, i.e., $p(H)=0$ if
$1-\left( 1-q\right) \beta H < 0$. Thus, when a classical
partition function is calculated, the integration limits on the
phase space are obtained from the condition $1-\left( 1-q\right)
\beta H\geq 0$.

The partition function (\ref{partition}) for the $N$ harmonic
oscillator system is
\begin{eqnarray}
Z_{q}&=&\int \prod_{n=1}^{N}{\mbox d}p_{n}{\mbox d}x_{n}\left[
1-(1-q)\beta \right.\nonumber \\
& & \times \left.\left(
\sum_{k=1}^{N}\frac{p_{k}^{2}}{2m}+\frac{1}{2}m\omega_n
^{2}x_{k}^{2}\right) \right] ^{1/(1-q)}\;.
\end{eqnarray}
To calculate this integral we introduce the variables:
$y_{n}=[(1-q)m\omega ^{2}\beta /2]^{1/2}x_{n}$ and
$y_{N+n}=[(1-q)\beta /(2m)]^{1/2}p_{n}$, where $n=1,2,3,...,N$. In
terms of these variables, $Z_{q}$ becomes
\begin{eqnarray}
Z_{q}&=&\left \{ \prod_{n=1}^{N}\left[ \frac{2}{(1-q)\omega_n
\beta
}\right] \right \} \nonumber \\
& &\times\int \prod_{n=1}^{2N}{\ \mbox d}y_{n}\left( 1-\sigma
\sum_{k=1}^{2N}y_{k}^{2}\right) ^{1/(1-q)}\; , \label{hyp}
\end{eqnarray}
with $\sigma ={\mbox {sign}}(1-q)$. For sufficiently large $N$
this integral diverges when $\sigma =-1$, {\it i.e.} when $q \geq
1+1/N$. This fact indicates that the Tsallis statistics with $q>1$
can not be employed for large number of subsystems. At this
moment, we restrict our discussion to the case $q\leq 1$.  By
using hyperspherical coordinates with $
u=(\sum_{n=1}^{2N}y_{n}^{2})^{1/2}$ and performing the integral
over the angular variables, we obtain
\begin{eqnarray}
Z_{q}&=&\left \{ \prod_{n=1}^{N}\left[ \frac{2}{(1-q)\omega_n
\beta
}\right] \right \}\frac{\Omega _{2N}}{2}\nonumber \\
& &\times \int_{0}^{1}{\mbox d}u\;u^{N-1}\left( 1-u\right)
^{1/(1-q)}\;.
\end{eqnarray}
By substituting the expression for the solid angle\cite{Grad1}
$\Omega _{2N}=2\pi ^{N}/\Gamma (N)$, and employing integral
representation of Euler beta function\cite{Grad1}, we verify that
\begin{eqnarray}
Z_{q}&=&\left \{ \prod_{n=1}^{N}\left[ \frac{2 \pi}{(1-q)\omega_n
\beta }\right] \right \}\frac{\Gamma \left( \frac{1}{1-q}+1\right)
}{\Gamma \left( \frac{1}{1-q}+1+N\right) }\nonumber \\
&=& \left[\left(\frac{2-q}{1-q}\right)_N\right]^{-1}
\prod_{n=1}^{N}\left[ \frac{2 \pi}{(1-q)\omega_n \beta }\right]
\;, \label{z}
\end{eqnarray}
where $(a)_n\equiv \prod_{k=0}^{n-1} (a+k)$ is the Pochhammer
symbol.

In the following analysis, one can consider $\omega_n =\omega$
without loss of generality. In fact, if someone is interested in
recovering the case  $\omega_n \neq \omega$, it is sufficient to
replace  $\omega^N$ with $\prod_{n=1}^{N}\omega_n$ in the
following equations.

To investigate the $(1-q)$ expansion and factorization
approximation we can employ any thermodynamical function. In this
work, we use the internal energy. As exposed below Eq.
(\ref{VINCULO1}), the generalized internal energy can be obtained
from the identity $U_q = F_q -T \partial F_q /\partial T$, so it
is given by
\begin{eqnarray}
U_q&=&\frac{N}{\beta}\; Z_{q}^{1-q}\nonumber \\
&=&\frac{N}{\beta}\left \{ \left [ \frac{2\pi }{(1-q)\omega \beta
}\right] ^{N}\left[\left(\frac{2-q}{1-q}\right)_N\right]^{-1}
\right \}^{1-q} . \label{UUqq}
\end{eqnarray}
In order to analyze the $(1-q)$ expansion, we consider a series of
Eq. (\ref{UUqq}) in powers of $(1-q)$,
\begin{eqnarray}
U_{q}&=& \frac{N}{\beta} \left[ \frac{2\pi }{\omega \beta }\right]
^{(1-q)N} \left [ 1- \frac{1}{2}(1-q)^{2}N(N+1)\right.\nonumber \\
&+&\left.\frac{1}{12}(1-q)^3N(N+1)(2N+1)+ ...\right ] .
\label{expand}
\end{eqnarray}
Since the $n$th term of this series essentially contains the
factor $[(1-q)N]^n$, the convergence of the series is improved if
$(1-q)N\ll 1$. In other words, the above expansion seems to be
reasonable for arbitrary $\beta ,m$ and $\omega $ only if
$(1-q)N\ll 1$.

Let us now analyze the possibility of factorization of the
partition function of a system with $N$ subsystems.
In other words, we want to understand when the partition
function can be well approximated by
\begin{eqnarray}
Z_{q}^{factor}=\prod_{i=1}^{N}Z_{q}^{(i)}\;,
\end{eqnarray}
where the upper index factor refers to the quantities obtained
with the factorization approach and $Z_{q}^{(i)}$ represents the
partition function of the subsystem labeled by $i$, and $N$ is the
number of subsystems. In the present study, each $ Z_{q}^{(i)}$ is
the partition function of one harmonic oscillator. Thus, the
partition function $Z_{q}^{factor}$ can be obtained from (\ref{z})
by using $N=1$, {\it i. e.}
\begin{equation}
Z_{q}^{factor}=\prod_{i=1}^{N}\frac{2\pi }{(2-q)\omega \beta
}=\left[ \frac{2\pi }{(2-q)\omega \beta }\right] ^{N} ,
\label{factorization-1}
\end{equation}
where we employed $Z^{(i)}=2\pi/((2-q)\omega \beta)$ and
$\omega_n=\omega$. The internal energy in this case is given by
\begin{eqnarray}\label{mala}
U_q^{factor}=\frac{N}{\beta}\left[ \frac{2\pi }{(2-q)\omega \beta
}\right]^{(1-q)N} \, .
\end{eqnarray}
After performing an expansion in powers of $(1-q)$,
Eq.(\ref{factorization-1}) can be written as
\begin{eqnarray}
U_{q}^{factor}&=&\frac{N}{\beta}\left[ \frac{2\pi }{(2-q)\omega
\beta }\right]^{(1-q)N}\nonumber \\
&=&\frac{N}{\beta}\left[ \frac{2\pi }{\omega \beta }\right]
^{(1-q)N} \nonumber \\
&\times&\left [ 1-(1-q)^2N-\frac{1}{2}(1-q)^3N+...\right ] .
\label{16}
\end{eqnarray}
The comparison between (\ref{expand}) and (\ref{16}) shows that
$(1-q)$ and $N$ dependences in $U_{q}$ and $U_{q}^{factor}$ come
out to be different. Similar differences are also present in the
other thermodynamical functions. Furthermore, it is important to
remark that the $N$ dependence on  Eqs. (\ref{UUqq}) and
(\ref{mala}) differs drastically for large $N$ (see Fig.(1)). In
addition, we also see in Fig.(1) that Eq.(\ref{UUqq}) and
Eq.(\ref{expand}) are closed for the values of $q$ and $N$ when
$(1-q)N\ll 1$ is verified. However, when this condition is not
satisfied, these equations lead to a significative deviation from
the exact result, Eq. (\ref{UUqq}) (see Fig.(1)). Furthermore, the
error due  to the approximation  is more pronounceable for the
factorization approximation than for the $(1-q)$ expansion. Thus,
$U_q^{factor}$ is not also an accurate approximation for $U_q$
when we consider an arbitrary $q$ and a very large $N$.

\section*{III- Exact versus approximate calculations and normalized constraints}

\begin{figure}
 \centering
 \DeclareGraphicsRule{ps}{eps}{*}{}
 \includegraphics*[width=10cm, height=14cm,trim=1cm 0cm 0cm 1cm]{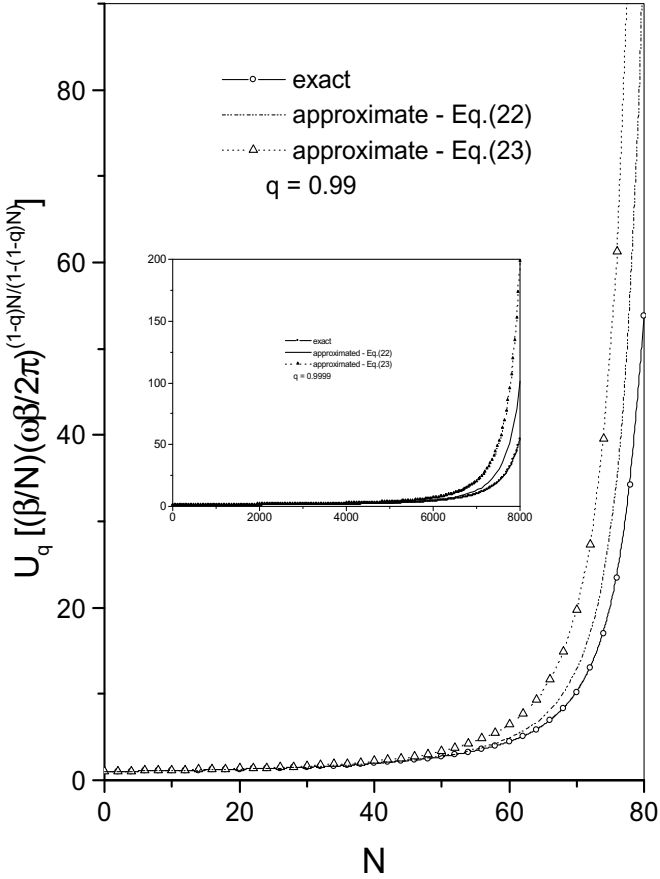}
\caption{In this figure, we show the differences among Eqs.
(\ref{UUqqnor}), (\ref{uqx}) and (\ref{uqw}) versus $N$ for
$q=0.99$ and $q=0.9999$.} \label{fig2}
\end{figure}

In this section, we analyze the differences among the
factorization approximation, $(1-q)$ expansion and the exact
calculation in the normalized approach of the Tsallis statistics.
The nonextensive canonical distribution $\rho_q^{nor}$, which
emerges from the maximum entropy principle\cite{ts88,ct91} and
subject to the normalized constraints\cite{NORMA}
\begin{eqnarray}
U_{q}^{nor}=\frac{\text{Tr}\left(\rho_{q}^{nor}\right)^{q}H}{\text{Tr}%
\left(\rho_{q}^{nor}\right)^{q}} \text{ \ \ \ \ \ \ and \ \ \ \ \
Tr}\rho_{q}^{nor}=1 \; ,\label{Uqnor}
\end{eqnarray}
is given by
\begin{eqnarray}
\rho_{q}^{nor}=\frac{1}{Z_{q}^{nor}} \left [1-(1-q)\frac{\beta
}{\text{Tr} \rho _{q}^{q}} \left( H-U_{q}^{nor}\right) \right
]^{1/(1-q)} \;\;\;\; , \label{mdnor}
\end{eqnarray}
where $Z_q^{nor}$ is the partition function
\begin{eqnarray}
Z_q={\mbox {Tr}}\left [1-(1-q)\frac{\beta }{\text{Tr} \rho
_{q}^{q}} \left( H-U_{q}^{nor}\right) \right ]^{1/(1-q)} \;\;\;\;
. \label{znor}
\end{eqnarray}
In general, the calculations in the normalized Tsallis statistics
may be obtained from the unnormalized one\cite{NORMA}. In this
way, employing Eqs. (\ref{Uqnor}) and (\ref{mdnor}) with the
Hamiltonian given by (\ref{1}) and taking into account Eq.
(\ref{z}), we obtain that
\begin{eqnarray}
U_{q,nor}&=&\frac{N}{\beta} \left\{ \left[ \frac{2\pi
}{(1-q)\omega \beta }\right]^N
\left[\left(\frac{2-q}{1-q}\right)_N\right]^{-1}\right\}
^{\frac{(1-q)}{1-(1-q)N}} \nonumber \\
& &\times \left[1+(1-q)N\right]^{\frac{1+(1-q)N}{1-(1-q)N}}
\label{UUqqnor}
\end{eqnarray}
for the exact case. In particular, Eq. (\ref{UUqqnor}) expressed
in terms of the effective temperature $T_{eff}={\mbox
{Tr}}(\rho_q^{nor})^q/\beta$ (for a discussion about effective
temperature and Lagrange parameters see \cite
{web,temperature,S1}) is $U_{q,nor}=N T_{eff}$.

Let us now evaluate the internal energy by considering $(1-q)$
expansion and the factorization approximations in order to perform
the analysis of these approximations in the normalized approach.
We start by expanding Eq. (\ref{UUqqnor}) in series of $(1-q)$ as
follows:
\begin{eqnarray}  \label{uqx}
U_{q,nor}&=& \frac{N}{\beta} \left[ \frac{2\pi }{\omega \beta
}\right]
^{\frac{(1-q)N}{1-(1-q)N}}\left[1+(1-q)N\right]^{\frac{1+(1-q)N}{1-(1-q)N}}
\nonumber \\ &\times& \left [ 1-\frac{1}{2}(1-q)^2N(N+1) \right.
\nonumber \\
&-& \left. \frac{1}{12}(1-q)^3N(N+1)(4N-1)+...\right ]  .
\end{eqnarray}
 The internal energy in the factorization
approach can be obtained employing similar calculation as the
previous one in Sec. II and it is given by
\begin{eqnarray} \label{uqw}
U_{q,nor}^{factor} &=&\frac{N}{\beta} \left[ \frac{2\pi
}{(2-q)\omega \beta }\right]^{\frac{(1-q)N}{1-(1-q)N}} \nonumber\\
&\times&
\left[1+(1-q)N\right]^{\frac{1+(1-q)N} {1-(1-q)N}} \nonumber \\
&=&\frac{N}{\beta} \left[ \frac{2\pi }{\omega \beta }\right]
^{\frac{(1-q)N}{1-(1-q)N}}\left[1+(1-q)N\right]^{\frac{1+(1-q)N}{1-(1-q)N}}
\nonumber \\ &\times& \left[1-(1-q)^2N +
\frac{1}{2}(1-q)^3N(2N-1)+... \right ]. \nonumber\\
\end{eqnarray}
The dependence on $N$ and $(1-q)$ in the expansions (\ref{uqx})
and (\ref{uqw}) is notoriously different. In fact, it essentially
remembers the previous one performed in Sec. II. Thus, the
conclusions about how useful (accurate) the $(1-q)$ expansion and
the factorization approximations are when describing the behavior
of the thermodynamical functions in normalized cases are similar
to the ones  presented in Sec. II, {\it{ i. e. }}, the
approximations seem reasonable only for $(1-q)N\ll 1$. The $N$
dependence on Eqs. (\ref{UUqqnor}), (\ref{uqx}) and (\ref{uqw}) is
also illustrated in Fig. (2). Moreover, as in the case of the
unnormalized Tsallis statistics, the $(1-q)$ expansion gives a
better agremment with the exact result than the factorization
approximation.

It is important to remark that similar results to those obtained
in Secs. II and III are also valid when $q > 1$. In this case, a
further care is necessary as the inequality $q < 1+ 1/N$ must be
obeyed so the partition function (\ref{hyp}) remains finite.

\section*{IV-Conclusion}

The previous discussions demonstrated that the $(1-q)$ expansion
and the factorization approximation are not useful for a system
with $N$ harmonic oscillators when someone employs an arbitrary
$q$ and a very large $N$. In particular, for systems  where
the number of oscillator is taken as the Avogradro number, i.e.,
$N \approx 10^{23}$,  $q \approx 1$ is implied, so it is
consistent with the discussion presented above. Similar
conclusions can be directly verified for the classical ideal gas,
since the calculation details related to it are similar to that
performed here for a set harmonic oscillators. We would like to
stress that the system discussed here is usually employed as a
first approximation in many contexts and consequently it has
received a special attention in the studies based on the
nonextensive Tsallis statistics.
 In conclusion, the results presented here strongly suggest that
$(1-q)$ expansion and factorization approximation are not good
approximating methods when $q$ is generic and $N$ is very large;
thus, we should be very careful when employing these
approximations.

\section*{Acknownlegement}
We gratefully acknowledge useful discussion with A. R. Plastino.
Also, we thank the CNPq and PRONEX for the financial support.


\end{document}